\documentclass{article}

\usepackage{amsmath,amssymb}

\newcommand{\Real}{\mathbb{R}}
\newcommand{\Comp}{\mathbb{C}}

\if@twoside \oddsidemargin 0.5cm
\else \oddsidemargin 0.5cm
\fi

\begin{document}

\centerline{\Large\bf SIGNATURE CHANGE AND CLIFFORD ALGEBRAS}

\vspace*{3ex}

\centerline{{\bf D. Miralles}$^{(1)}$,
{\bf J. M. Parra}$^{(1)}$
{\bf and\/} 
{\bf J. Vaz, Jr.}$^{(2)}$}

\vspace*{1ex}

\begin{center}
(1) Departament de F\'{\i}sica Fonamental\\
Universitat de Barcelona\\
Diagonal 647, CP 08028, Barcelona, Catalonia, Spain\\
\vspace{.5ex}
(2) Departamento de Matem\'atica Aplicada\\
Universidade Estadual de Campinas\\
CP 6065, 13081-970 Campinas, SP, Brazil 
\end{center}

\begin{abstract}
Given the real Clifford algebra of a quadratic space with a
given signature, we define a new product in this structure
such that it simulates the Clifford product of a quadratic
space with another signature different from the
original one. Among the possible applications of this
new product, we use it in order to write 
the minkowskian Dirac equation over the euclidean spacetime 
and to define a new duality operation in terms of 
which one can find self-dual and anti-self-dual 
solutions of gauge fields over Minkowski spacetime 
analogous to the ones over Euclidean spacetime and
without needing to complexify the original real algebra.
\end{abstract}

\section{Introduction}

Clifford algebras (CA) are very important in theoretical and 
mathematical physics. It is almost impossible to list all its
applications but the interested reader can found some of them 
in references \cite{CAMP}. Now, in order to define 
the CA of a given vector space $V$ one need to endow 
$V$ with a (in general symmetric) bilinear form $g_{p,q}$. 
One interesting fact is that the structure 
of the CA depends not only on the dimension of $V$ but also 
on the signature of $g$. In another words, real Clifford algebras 
$Cl_{p,q}$ and $Cl_{p^\prime,q^\prime}$ ($p+q = 
p^\prime+q^\prime = n$) 
are in general not isomorphic, that is, when we change the
signature we get in general different Clifford algebras. 
For example, the Clifford algebras $Cl_{1,3}$ and 
$Cl_{3,1}$ are not isomorphic -- indeed, in terms of 
matrix algebras, the
former is isomorphic to the algebra of $2 \times 2$ quaternionic
matrices while the latter is isomorphic to the algebra of 
$4 \times 4$ real matrices. 

Changing the signature of a given space may sound in 
principle an artificial process but undoubtly it is a 
very important thing in modern physics. For example, the 
euclidean formulation of field theories is a fundamental tool 
in modern physics \cite{GJ}. Indeed, sometimes it seems to be even
crucial, as in the theory of instantons, in finite 
temperature field theory and in lattice gauge theory. 
Going from an euclidean to a minkowskian theory or vice-versa
involves changing the signature of the metric over the
spacetime and in general a minkowskian theory 
is transformed in an euclidean theory by analytical continuation, 
that is, by making $t \rightarrow i t$. 
The interpretation 
of making $t \rightarrow i t$ is not trivial -- in relation 
to this, see \cite{NW} (and references therein) where 
it was interpreted as a rotation in a five dimensional 
spacetime. 

Despite the ingenuity of an approach 
like \cite{NW} in interpreting $t \rightarrow i t$ as a 
rotation in a five dimensional space, we believe it is unsatisfactory 
since the use of an additional time coordinate in spacetime appears
to us to be meaningless. Indeed it would be much more
satisfactory if one could find an approach to describe 
the signature change where there is no need to 
introduce extra dimensions. In the case where the problem 
involves CA the situation is even more problematic since 
the signature changed CA and the original one may not 
be isomorphic. A possible way to overcome this situation 
is to complexify the original real CA since the structure of 
complex CA depends only on the dimension of $V$, but this 
approach can also be seen as a result of introducing an 
extra dimension (see below).

The objective of this paper is to introduce an 
{\it algebraic\/} approach to the change of signature 
in CA where there is {\it no need to introduce
any extra dimension} to describe it.  
The signature change appears as a transformation on the 
algebraic structure underlying the theory. The idea is
to propose an operation that ``simulates'' the product properties of
the signature changed space in terms of the original space 
or vice-versa. The particular case we have in mind is the
one involving the four dimensional spacetime, where we
introduce an operation ``simulating'' the properties of 
minkowski spacetime in terms of an euclidean spacetime 
and vice-versa. This operation will be called ``vee product'' 
(since it will be denoted by a $\vee$ in order to distinguish 
it from the usual Clifford product that will be denoted by 
juxtaposition). 
The advantage of this approach is obvious 
since we can retain the ``physics'' (the minkowskian properties) 
in a suitable mathematical world (the euclidean spacetime).  
The fact that we can define the vee product in terms of the 
Clifford product means that we can describe the minkowskian 
properties in terms of euclidean spacetime and vice-versa. 

Our approach has been inspired by the work of 
Lounesto \cite{Lounesto}. But there are some differences 
in the method and interpretation which the reader can 
compare. Moreover, Lounesto only discussed some problems 
involving the case of signature change corresponding 
to opposite signatures, that is, $(p,q)$ and $(q,p)$. 
Cases like $(p,q)$ and $(p+q,0)$ have not been considered 
by Lounesto and this is the case we have when 
considering minkowskian and euclidean spacetimes. 
Some applications of this are discussed here -- and
some others can be found in \cite{Pezzaglia}.

As an application, we are first interested in this paper 
in studying the Dirac equation. First of all, the version 
of Dirac equation we obtain by making $t \rightarrow i t$ -- 
we shall call it the euclidean Dirac equation -- has 
physical properties that are obviously different from the 
original Dirac equation -- which we shall call the minkowskian 
Dirac equation. The question we want to address in this 
context is if it is possible to obtain a new equation 
that exhibit the same physical properties of the original 
equation. More specifically, our idea is to write
the minkowskian Dirac equation in the euclidean spacetime, 
which should obviously be different from an euclidean 
Dirac equation in an euclidean spacetime. Minkowskian 
and euclidean spacetimes are different worlds, both 
mathematically and physically speaking, and we want to 
simulate the minkowskian scenario in an euclidean world. 
The idea is to write an equation in euclidean spacetime 
in terms of the vee product such that it is equivalent 
to the original equation in terms of the Clifford product 
in Minkowski spacetime. Of course the Dirac equation in terms of 
the vee product is expected to be different from the euclidean 
Dirac equation.

In order to introduce our approach we need first of all 
to take some care. Dirac equation is in general formulated
in terms of $4 \times 4$ complex matrices - the gamma matrices - 
obeying the relation $\Gamma_\mu \Gamma_\nu + \Gamma_\nu
\Gamma_\mu = 2 g_{\mu\nu}$, where $g_{\mu\nu} = {\rm diag}(1,1,1,1)$ 
in the euclidean case and $g_{\mu\nu} = {\rm diag}(1,-1,-1,-1)$ in 
the minkowskian case. This algebra is the matrix representation 
of the complex Clifford algebra 
$Cl_{\scriptscriptstyle \Comp}(4)$\cite{Porteous}. On the 
other hand, this algebra is the complexification of the 
real algebras $Cl_{1,3}$ and $Cl_{4,0}$ 
associated with the minkowskian and euclidean spacetimes, 
respectively, that is, 
$Cl_{\scriptscriptstyle \Comp}(4) 
\simeq \Comp \otimes Cl_{1,3} 
\simeq \Comp \otimes Cl_{4,0}$. 
Moreover, $Cl_{\scriptscriptstyle \Comp}(4)$ is also 
isomorphic to the real algebra
$Cl_{4,1}$, while the real algebras 
$Cl_{1,3}$ and $Cl_{4,0}$ - which 
are also isomorphic\footnote{However $Cl_{3,1}$ and $Cl_{4,0}$ 
are not isomorphic, and we shall see how to consider this case
also. Anyway, it is not the isomorphism between $Cl_{1,3}$ and
$Cl_{4,0}$ that matters in this discussion.} - are
isomorphic to the even 
subalgebra of $Cl_{4,1}$ \cite{Rod}. 
All these facts show that when we complexify the 
real algebra $Cl_{1,3}$ getting 
$\Comp \otimes Cl_{1,3} \simeq 
Cl_{\scriptscriptstyle \Comp }(4) \simeq 
Cl_{4,1}$ we are introducing an extra 
dimension to the minkowskian spacetime such that
the extra dimension is of the type of euclidean time. 
In the same way, when we complexify the real algebra
$Cl_{4,0}$ getting $\Comp \otimes
Cl_{4,0} \simeq Cl_{\scriptscriptstyle \Comp}(4)
\simeq Cl_{4,1}$ we are introducing an extra
dimension to the euclidean spacetime such that the
extra dimension is of the type of minkowskian time. 

Now, our idea is to not use any extra dimension, and 
{\it one certain way to do this\/} is to avoid using any
of these complexified structures. This can be achieved
using the real formulation of Dirac theory due to 
Hestenes \cite{Hest1}. One can easily formulate the
Dirac theory in terms of the Clifford algebra 
$Cl_{1,3}$, which is isomorphic to the 
algebra of $2 \times 2$ quaternionic matrices. 
A Dirac spinor in this way is represented by a 
pair of quaternions, but rather we prefer to use
the form of Dirac equation called Dirac-Hestenes
equation in terms of the so called Dirac-Hestenes
spinor \cite{Rod}. The reason for our choice is 
simple. The Dirac-Hestenes spinor is represented 
by an element of the even subalgebra $Cl_{1,3}^+$ 
while a Dirac spinor is an element of an ideal of 
$Cl_{1,3}$. Both formulations are 
equivalent \cite{IJTP} but if we use the former one
we avoid the problem of considering the transformation
between different ideals (in fact left and right ideals), 
which will happen in the 
latter case, and in order to not do unnecessary work --
and even hidden some fundamental facts -- 
we prefer to use the Dirac-Hestenes equation.

As another application, we discuss the problem of
finding self-dual and anti-self-dual solutions of gauge 
fields. Since the group being abelian or not is irrelevant for 
this matter we shall restrict our attention to the abelian 
case. As is well-known, in the Minkowski spacetime there does not
exist (real) solutions to the problem $* F = \pm F$, 
where $F$ is the 2-form representing the electromagnetic 
field and $*$ is the Hodge star operator. However, in an
euclidean spacetime this problem has solutions and they 
are given by $E = \pm B$, where $E$ and $B$ are the 
electric and magnetic components of $F$. Now, using the 
operation we discussed above, we can define a new Hodge-like
operator on Minkowski spacetime such that in relation to 
this new operator we have self-dual and anti-self-dual solutions
for that problem on Minkowski spacetime. Moreover, using
this Hodge-like operator we can completely simulate an 
euclidean metric while still working on Minkowski spacetime. 
We also show that the relation between those Hodge-like 
operators is given by the parity operation. 

We organized this paper as follows. In section 2 we 
briefly discuss the Clifford algebras. In section 3
we discuss the transformation from euclidean to 
minkowskian spacetimes and vice-versa from an algebraic 
point of view. Our idea is to define a new Clifford
product in terms of the original Clifford product 
that defines the algebra we are working. This new 
product simulates the 
algebra of Minkowski spacetime inside the algebra 
of the euclidean spacetime. Then in section 4 we 
discuss the minkowskian Dirac equation over the
euclidean spacetime using our approach. Of course
that one could be interested in the other way, that
is, to simulate an euclidean world inside a minkowskian
spacetime. Indeed we can consider any other possibility, 
as we discuss in section 5.
Finally in section 6 we discuss how to obtain
solutions corresponding to self-dual and anti-self-dual 
gauge fields and the definition of a Hodge-like operation 
with many interesting properties.

\section{Mathematical Preliminaries}

There are many different ways to define Clifford 
algebras \cite{citas}, each of them
emphasizing different aspects. Our approach has been choosen due to 
the direct introduction to Clifford product \cite{r-57} which 
will be fundamental in the next sections.

Let $\{{\bf e}_{1},\ldots,{\bf e}_{n}\}$ be an orthonormal basis
for $\Real^{p,q}$, where $\Real^{p,q}$ is a real vector space of
dimension $n=p+q$ endowed with an interior product
$g:\Real^{p,q}\times \Real^{p,q} \rightarrow \Real$. Writing the 
quadratic form $\sum_{i=1}^{n}\sum_{j=1}^{n} g_{ij} x_{i} x_{j}$ 
as the square of the linear expression $\sum_{i=1}^{n} x_{i}{\bf e}_{i}$
and assuming the distributive property we
obtain the well-known expression for the Clifford algebra
$Cl(\Real^{p,q},g)\equiv Cl_{p,q}$,
\begin{equation}
{\bf e}_{i}{\bf e}_{j}+{\bf e}_{j}{\bf e}_{i}=2g_{ij}
\end{equation}
where $g_{ij}$ are the metric components. This defines the Clifford
product, which has been denoted by juxtaposition.

There is a product $\wedge$, called the exterior product, underlying the 
Clifford algebra. It is an associative, bilinear and skew-symmetric
product of vectors. Furthermore, by applying it to our orthogonal basis
we can construct a new vector space $\Lambda^{2}(\Real^{p,q})$ whose 
elements are called bivectors, i.e., $\wedge :\Real^{p,q} \times \Real^{p,q}
\rightarrow \Lambda^{2}(\Real^{p,q})$. The skew-symmetric property allows us
to extend the definition to $\Lambda^{n}(\Real^{p,q})$. In general
$\wedge :\Lambda^{k}(\Real^{p,q})\times\Lambda^{l}(\Real^{p,q})\rightarrow
\Lambda^{k+l}(\Real^{p,q})$. \\
If $\{{\bf e}_{1},\ldots,{\bf e}_{n}\}$
is a basis of $\Real^{p,q}$ then $1$ and the Clifford products
${\bf e}_{i_{1}}\cdots{\bf e}_{i_{k}}$, $(1\leq i_{1}<i_{2}< \ldots 
<i_{k} \leq n)$ will establish a basis for $Cl_{p,q}$ which has 
dimension $2^{n}$. If $\{{\bf e}_{1},\ldots,{\bf e}_{n}\}$ is an 
orthogonal basis then ${\bf e}_{1}\cdots{\bf e}_{n}=
{\bf e}_{1}\wedge \cdots \wedge {\bf e}_{n}$, which is usually called volume
element. It follows that $Cl_{p,q}$ and $\Lambda(\Real^{p,q})
= \oplus_{k=0}^n \Lambda^k(\Real^{p,q})$ are 
isomorphic as vector spaces. Therefore, a general
element $A\in Cl_{p,q}$ takes the form
\begin{equation}
A=A_{0}+A_{1}+\ldots+A_{n}
\end{equation}
where $A_{r}$, called an r-vector, belongs to $\Lambda^{r}(\Real^{p,q}) \subset Cl_{p,q}$,
$(r=0,1,\ldots,n)$. More explicitly,
\begin{equation}
A=a_{0}+a^{i}{\bf e}_{i}+a^{ij} {\bf e}_{ij}+\cdots+a^{1 \ldots n} 
{\bf e}_{1 \ldots n}
\end{equation}
It is convenient to define a projector $\langle \, \rangle_{r}$ 
as $\langle \, \rangle_{r}:\Lambda(V) \rightarrow
\Lambda^{r}(V)$, i.e., $\langle A \rangle_{r}=A_{r}$. \\
An important property is that Clifford algebra is a $Z_{2}$-graded
algebra, i.e., we can divide it into even ($Cl^{+}_{p,q}$) and odd
($Cl^{-}_{p,q}$) grades. $Cl^{+}_{p,q}$ is a sub-algebra of $Cl_{p,q}$
called even sub-algebra. In our example, $Cl^{+}_{3,0}=\{1,{\bf e}_{12},
{\bf e}_{13},{\bf e}_{23}\}$ and $Cl^{-}_{3,0}=\{{\bf e}_{1},{\bf e}_{2},
{\bf e}_{3},{\bf e}_{123}\}$. Some important identities that we will use later are 
($a \in \Real^{p,q}$, $B \in \Lambda^{r}(\Real^{p,q}) \subset Cl_{p,q}$):
\begin{equation}
aB=a\cdot B+a \wedge B
\end{equation}
\begin{equation}
a \wedge B\equiv \langle aB \rangle_{r+1} =\frac{1}{2}(aB+(-1)^{r}Ba)
\end{equation}
\begin{equation}
a \cdot B \equiv \langle aB \rangle_{r-1} =\frac{1}{2}(aB-(-1)^{r}Ba)
\end{equation}

\section{The Vee Product}

Let $V$ be a vector space of dimension $n=4$. We have five different
Clifford algebras depending on the signature: $Cl_{4,0}$, $Cl_{3,1}$, 
$Cl_{2,2}$, $Cl_{1,3}$ and $Cl_{0,4}$. With the (possible) exception of
$Cl_{2,2}$ 
the importance of the others in modern physics is more than obvious.

First we shall consider the case involving the algebras $Cl_{1,3}$ 
and $Cl_{4,0}$. 
Let $A, B  \in Cl_{4,0}$ and $AB$ be its Clifford product. Now we define
a new product, which we call a vee product, $A \vee B$  simulating
the $Cl_{1,3}$ Clifford product in $Cl_{4,0}$. After the selection in 
$\Real^{4,0}$ of an arbitrary unit vector ${\bf e}_{0}$
to represent the {\it fourth dimension} and the completion of the basis
with three other orthonormal vectors ${\bf e}_{i}$, we define for 
${\bf u}, {\bf v} \in \Real^{4,0}$

\begin{equation}
{\bf u} \vee {\bf v}:=(-1)[{\bf v} {\bf u}-
2({\bf v} \cdot {\bf e}_{0}) ({\bf e}_{0} \cdot {\bf u})]
\end{equation}
Using this product and the $Cl_{4,0}$ standard basis it is easy
to prove that ${\bf e}_{0} \vee {\bf e}_{0}=1$ and ${\bf e}_{i}
\vee {\bf e}_{i}=-1$ ($i = 1,2,3$), while 
${\bf e}_\mu^2 = {\bf e}_\mu{\bf e}_\mu  = 1$ ($\mu = 0,1,2,3$). 
Moreover, for  ${\bf u},{\bf w}
\in \Real^{4,0}$, 
\begin{equation}
{\bf u}{\bf w}+{\bf w}{\bf u}=2{\bf u} \cdot {\bf w}=
2(u_{0}w_{0}+u_{1}w_{1}+u_{2}w_{2}+u_{3}w_{3})
\end{equation}
but if we use the vee product then
\begin{eqnarray}
{\bf u} \vee {\bf w}+{\bf w} \vee {\bf u} & =
& -{\bf w}{\bf u}+2({\bf w} \cdot {\bf e}_{0})({\bf e}_0 
\cdot {\bf u})-{\bf u}{\bf w}+2({\bf u} \cdot {\bf e}_{0})
({\bf e}_0 \cdot {\bf w}) \nonumber \\
& = &  2(u_{0}w_{0}-u_{1}w_{1}-u_{2}w_{2}-u_{3}w_{3})\nonumber
\end{eqnarray}
A little bit more general case is when one has a vector and a
$k$-graded element, i.e., ${\bf v}$ and $B_{k}$
\begin{equation}
B_{k} \vee {\bf v}= (-1)^{k}[{\bf v}B_{k}-2({\bf v} \cdot {\bf e}_{0})
({\bf e}_{0} \cdot B_{k})]
\end{equation}
\begin{equation}
{\bf v} \vee B_{k}= (-1)^{k}[B_{k}{\bf v}-2(B_{k} \cdot {\bf e}_{0})
({\bf e}_{0} \cdot {\bf v})]
\end{equation}
Now with the help of (6) one can see that $\vee $ is associative,  
that is, ${\bf v}\vee({\bf u}\vee{\bf w}) = ({\bf v}\vee{\bf u})\vee
{\bf w}$.  
Moreover,
the vee product preserves the multivectorial structure since 
\begin{equation}
\frac{1}{2}[{\bf u},{\bf v}]_\vee=
\frac{1}{2}({\bf u}\vee {\bf v}-{\bf v}\vee {\bf u})= 
\frac{1}{2}(-{\bf v}{\bf u}+2 u_0 v_0+{\bf u}{\bf v}-2 u_0 v_0)=
\frac{1}{2}({\bf u}{\bf v}-{\bf v}{\bf u})=
\frac{1}{2}[{\bf u},{\bf v}]={\bf u}\wedge {\bf v} \nonumber
\end{equation}
Finally, we can generalize those expression as
\begin{equation}
A_{l} \vee B_{k}=(-1)^{kl}[B_{k}A_{l}-2(B_{k} \cdot {\bf e}_0)
({\bf e}_{0} \cdot A_{l})]
\end{equation} 

\section{Dirac equation and vee product}

In this section we come to consider the results recently
introduced and their applications to Dirac equation.
We shall use Dirac-Hestenes 
equation \cite{Rod,Hest1,IJTP,l-93,pa-92} as
discussed in the introduction. 
As starting point for our analysis we take the Dirac-Hestenes
equation in $Cl_{1,3}$
\begin{equation}
\nabla\psi\gamma_{21}-m\psi\gamma_{0}=0
\end{equation}
where $\nabla$ denotes the Dirac operator, that is, 
$$\nabla=\gamma_{0}\partial_{0}-\gamma_{1}\partial_{1}-
\gamma_{2}\partial_{2}-\gamma_{3}\partial_{3}$$
and $\gamma_{\mu}$ ($\mu=0,1,2,3$) are interpreted as 
vectors in $Cl_{1,3}$ and
$\psi = \psi(x) \in Cl^{+}_{1,3}, \forall x \in M $,
where $M$ is the minkowskian manifold.
Now let us to ask a question: How can one simulate the minkowskian 
Dirac equation in an euclidean formulation? The answer is given 
for the vee product. \\

Firstly, multiplying on the right for $\gamma_{12}$ we can write 
(12) as 
\begin{equation}
\nabla\psi-m\psi\gamma_{012}=0
\end{equation}
Considering $\psi \in
Cl^{+}_{4,0}, \forall x \in M$ and using ${\bf e}$-notation for the
$Cl_{4,0}$ elements, 
the Dirac equation (13) can be written in the euclidean spacetime 
using the $\vee$ product as
\begin{equation}
\nabla \vee \psi-m\psi \vee {\bf e}_{012}=0
\end{equation}
where $\nabla = {\bf e}^\mu\partial_\mu$ with ${\bf e}^\mu =
{\bf e}_\mu$. Note that ${\bf e}_{012} = {\bf e}_0 {\bf e}_1
{\bf e}_2 = {\bf e}_0 \vee {\bf e}_1 \vee {\bf e}_2$. \\
Let us see how this equation appears in terms of the original 
Clifford product in euclidean spacetime. First 
we split the Dirac operator in temporal and space parts,
$$
\nabla\vee \psi={\bf e}_{0} \vee \partial_{0}\psi + 
{\bf e}_{i} \vee \partial_{i}\psi
$$
Working on the temporal part we have
\begin{eqnarray*}
{\bf e}_{0} \vee \partial_{0}\psi & = &
\partial_{0}\psi{\bf e}_{0}-2(\partial_{0}\psi \cdot {\bf e}_{0})
({\bf e}_{0} \cdot {\bf e}_{0}) \\
& = & \partial_{0}\psi{\bf e}_{0}-2[\frac{1}{2}(\partial_{0}\psi{\bf e}_{0}-
{\bf e}_{0}\partial_{0}\psi)]={\bf e}_{0}\partial_{0}\psi
\end{eqnarray*}
where we have used $\partial_{0}\psi \cdot {\bf e}_{0}=
\frac{1}{2}(\partial_{0}\psi{\bf e}_{0}-{\bf e}_{0}\partial_{0}\psi)$
For the space part we have
$$
{\bf e}_{i} \vee \partial_{i}\psi=
\partial_{i}\psi{\bf e}_{i}-2[(\partial_{i}\psi) \cdot
({\bf e}_{0} \cdot {\bf e}_{i})]=\partial_{i}\psi{\bf e}_{i}
$$
Therefore
\begin{equation}
\nabla \vee \psi={\bf e}_{0}\partial_{0}\psi+
\partial_{i}\psi{\bf e}_{i}
\end{equation}
It is easy to see that 
\begin{equation}
\nabla \vee \nabla \vee \psi=\square_{M} \psi
\end{equation}
where $\square_{M} = \partial_0^2 -\sum_{i=1}^3 \partial_i^2$, 
while $\nabla^2\psi = \square_E \psi$ with $\square_E = \partial_0^2 
+ \sum_{i=1}^3 \partial_i^2$. \\

The massive term in (13) will be
\begin{equation}
m\psi \vee {\bf e}_{012}=m{\bf e}_{12}\psi
{\bf e}_{0}
\end{equation}
Now we can write an equation simulating the minkowskian Dirac
equation in $Cl_{4,0}$:
\begin{equation}
{\bf e}_{0}\partial_{0}\psi+
\partial_{i}\psi{\bf e}_{i}-m{\bf e}_{12}\psi{\bf e}_{0}=0 
\end{equation}
If one considers a charged fermion field $\psi$ in interaction with 
the electromagnetic field $A$ we will add to the Dirac equation (13)
the term $ eA\psi \gamma_{12} $ where $A \in \Real^{1,3}$, applying the 
vee product we get

\begin{equation}
A \vee \psi \vee {\bf e}_{12}={\bf e}_{12}(\psi A-2(\psi \cdot {\bf e}_{0}) A_{0})
\end{equation} 
It is important to remark that {\it we have never changed the algebra} --
we have working with $Cl_{4,0}$.
It can be proved that solutions of (13) are solutions of (18) too.
For a general $\psi \in Cl_{1,3}^{+}, \forall x \in M$, we get in (13)
eight coupled 
differential equations; transforming this original $\psi$ solution to its
$Cl_{4,0}^{+}$ version and checking it with (18) we recover exactly the
same coupled system.

\section{The General Case}

Lounesto has studied \cite{l-93} the case involving 
the opposite signatures $(+---)$ and $(-+++)$. 
For $a,b \in Cl_{1,3}$ the tilt transformation, based in the even-odd
decomposition $Cl_{1,3}=Cl_{1,3}^+ \oplus Cl_{1,3}^-$ first emphasized
by Clifford \cite{cl-78}, was defined as
$$ 
\underbrace{ab}_{Cl_{1,3} \, product} \rightarrow 
\underbrace{b_{+}a_{+}+b_{+}a_{-}
+b_{-}a_{+}-b_{-}a_{-}}_{Cl_{3,1} \, product} 
$$
where the subscripts minus and plus are the odd and even parts.
We reinterpret this transformation as a new product $\vee_t$ given by 
\begin{equation}
A_{l} \vee_{t(3,1)} B_{k}=(-1)^{kl}B_{k}A_{l}
\end{equation}
The meaning of this expression is that we are defining a 
mapping from an algebra $Cl$ to its opposite
algebra $Cl^{\rm opp}$, and it is not difficult
to show that $Cl_{p,q}^{\rm opp} = Cl_{q,p}$. \\
In order to consider the general case, we consider 
the vee and tilt
products. There are indeed some similarities
\begin{equation}
{\rm Vee \, product:} \qquad \, \, A_{l} \vee
B_{k}=(-1)^{kl}[B_{k}A_{l}-2(B_{k} \cdot {\bf e}_0)
({\bf e}_{0} \cdot A_{l})]
\end{equation} 
\begin{equation}
{\rm Tilt \, product:} \qquad \, \,A_{l} \vee B_{k}=(-1)^{kl}B_{k}A_{l}
\end{equation}
It is interesting to see which is the purpose of the term with
the temporal component ${\bf e}_{0}$. The vee product simulates
the change of signature as 
$$\underbrace{(++++)}_{Cl_{4,0}}
\rightarrow\underbrace{(+---)}_{Cl_{1,3}}$$ 
where these signs
correspond to the square of the canonical basis elements. All
the squares change except for the temporal component. If we set
up the inverse problem $(+---)\rightarrow(++++)$, 
the general expression for the corresponding vee
product will be the same. This is very welcome since, for
example, we can get back the euclidean Dirac equation from 
the minkowskian one as in (18) just by using again the
$\vee$ product. The same holds in the opposite direction, 
of course. \\
For the tilt product, 
$$\underbrace{(+---)}_{Cl_{1,3}}
\rightarrow\underbrace{(-+++)}_{Cl_{3,1}}$$ 
we change all
the squares and now we do not need to subtract any temporal
term. Again the opposite problem $(-+++)\rightarrow(+---)$  
keeps up the same form.
For the most interesting cases in physics we have
$$\begin{array}{cc}
(++++)&\leftrightarrow(+---) \\
\updownarrow &\updownarrow \\
(----)&\leftrightarrow(-+++)
\end{array}$$
As we know the change of signature in the same row is done using 
the vee product and between rows of the same column using the tilt
product. 
Now if we want to do the change in a {\it diagonal} way, i.e.
$(++++)\leftrightarrow(-+++)$ then we will need to compose vee
and tilt product. 
All this products can be extended to anothers signatures. For example
$$\begin{array}{ccccc}
(++++)&\leftrightarrow(+---)&\leftrightarrow(--++)&\leftrightarrow(---+)&\leftrightarrow(----)\\  
\updownarrow &\updownarrow &\updownarrow &\updownarrow
&\updownarrow \\
(----)&\leftrightarrow(-+++)&\leftrightarrow(++--)&\leftrightarrow(+++-)&\leftrightarrow(++++)
\end{array}$$
If we want to change the signature along the same row we
only need to know the canonical basis element 
${\bf e}_{\mu}$ which doesn't change its square, then
\begin{equation}
A_{l} \vee B_{k}=(-1)^{kl}[B_{k}A_{l}-2(B_{k} \cdot {\bf e}_{\mu})
({\bf e}^{\mu} \cdot A_{l})]
\end{equation}
where ${\mu}$'s are {\it not\/} summed.
And for the change between rows in the same column
\begin{equation}
A_{l} \vee_t B_{k}=(-1)^{kl}B_{k}A_{l}
\end{equation}
As one can easily see there is no condition limiting this work to four
dimensions, so our scheme holds in any dimension. 

\section{Other Applications}

While in our discussion of the Dirac equation we were interested
in simulating ``minkowskian properties'' in an euclidean spacetime,
in this section we are interested in simulating ``euclidean
properties'' in Minkowski spacetime. Therefore in relation to
the notation we are interested in simulating the product of
$Cl_{4,0}$ in terms of the generators $\{\gamma_\mu\}$
($\mu = 0,1,2,3$) of $Cl_{1,3}$.

\subsection{The Hodge Star Operator}

The Hodge star operator $\ast$ in Minkowski spacetime can be
written using $Cl_{1,3}$ as
\begin{equation}
\label{eq.6.1}
\ast \Phi = \tilde{\Phi} \gamma_5 ,
\end{equation}
where $\Phi \in Cl_{1,3}$ is a multivector field and
$\gamma_5 = \gamma_0 \gamma_1 \gamma_2 \gamma_3$ and by
the tilde we denoted the reversion operation such that
\begin{equation}
\tilde{A}_k = (-1)^{k(k-1)/2} A_k
\end{equation}
for $A_k \in \Lambda^k$.

Using the vee product we can write the Hodge star operator
$\star$ corresponding to euclidean spacetime in terms of
the algebra $Cl_{1,3}$ of Minkowski spacetime
as\footnote{Note that the Hodge operator in euclidean spacetime
is denoted by an asterisk while the one in Minkowski spacetime
is denoted by a star.}
\begin{equation}
\label{eq.6.2}
\star \Phi = \tilde{\Phi}\vee \gamma_5 .
\end{equation}
It is easy to see that $\gamma_0\vee\gamma_1\vee
\gamma_2\vee\gamma_3 = \gamma_0\gamma_1\gamma_2\gamma_3$
and that $\gamma_5\vee\gamma_5 = 1$ while $\gamma_5\gamma_5 = -1$.
In order to rewrite the above expression using the definition
of the vee product it is convenient to split
$\Phi$ into even and odd parts, that is,
$\Phi = \Phi_+ + \Phi_-$ where $\Phi_\pm = \pm \hat{\Phi}_\pm$,
where by the hat we denoted the graded involution, that is,
\begin{equation}
\hat{A}_k = (-1)^k A_k ,
\end{equation}
where $A_k \in \Lambda^k$. Now we can write using the
definition of the vee product that
\begin{equation}
\label{eq.6.3}
\begin{split}
\Phi\vee\gamma_5 & = \Phi_+ \vee \gamma_5 + \Phi_- \vee \gamma_5 \\
& = \gamma_5 \Phi_+  + 2\gamma_{123}(\gamma_0\cdot\Phi_+) +
\gamma_5\Phi_- + 2\gamma_{123}(\gamma_0\cdot \Phi_-) ,
\end{split}
\end{equation}
and
\begin{equation}
\label{eq.6.4}
\Phi_+\vee \gamma_5 = \gamma_5 \gamma_0 \Phi_+ \gamma_0 , \qquad
\Phi_-\vee \gamma_5 = -\gamma_5 \gamma_0 \Phi_- \gamma_0 .
\end{equation}
We have therefore
\begin{equation}
\label{eq.6.5}
\Phi\vee\gamma_5 = \gamma_5 \gamma_0 \hat{\Phi} \gamma_0 ,
\end{equation}
and the euclidean Hodge star operator $\star$ can be
written as
\begin{equation}
\label{eq.6.6}
\star \Phi = \gamma_5 \gamma_0 \overline{\Phi}\gamma_0 ,
\end{equation}
where $\overline{\Phi} = \widehat{\tilde{\Phi}} =
\widetilde{\hat{\Phi}}$. Moreover, since the parity
operation $\mathcal{P}$ in $Cl_{1,3}$ is given by \cite{Hest1,l-93}
\begin{equation}
\label{eq.6.6.1}
\mathcal{P}(\Phi) = \gamma_0 \Phi \gamma_0 ,
\end{equation}
and if we rewrite (\ref{eq.6.6}) as
\begin{equation}
\label{eq.6.6.2}
\star \Phi = -\gamma_0 \widetilde{\Phi} \gamma_5 \gamma_0 ,
\end{equation}
we have that
\begin{equation}
\label{eq.6.6.3}
\star \Phi = - \mathcal{P}(\ast \Phi) .
\end{equation}
This expression clearly shows that $\star$ is written only
in terms of operations on Minkowski spacetime.

\subsection{Differential and Codifferential Operators}

In Minkowski spacetime the differential ${\rm d}$ and
codifferential $\delta$ operators
can be written in terms of Dirac operator $\nabla$
as \cite{Hest1}
\begin{equation}
\label{eq.6.7}
{\rm d}\Phi = \frac{1}{2}\left( \nabla\Phi +
\hat{\Phi}\overleftarrow{\nabla}\right) , \qquad
\delta\Phi = \frac{1}{2}\left( \nabla\Phi -
\hat{\Phi}\overleftarrow{\nabla}\right) ,
\end{equation}
and such that $\nabla = {\rm d} + \delta$, where $\nabla =
\gamma^\mu\partial_\mu$ and the right action of the Dirac operator
denoted by $\overleftarrow{\nabla}$ is defined
as $\Phi\overleftarrow{\nabla} = (\partial_\mu \Phi)\gamma^\mu$.
With the above definition we have that $\delta = \ast {\rm d} \ast$,
where $\ast$ is the Hodge star operator in Minkowski
spacetime\footnote{One can find different definitions for the
codifferential operator $\delta$ but this is completely
irrelevant for our purpose that is to give examples of
applications of our method.}.

Now we want to write the differential and codifferential operators
corresponding to the case of euclidean spacetime using the algebra
of Minkowski spacetime. Let us denote these operators by $\check{\rm d}$
and $\check{\delta}$, respectively, the check being used to distinguish
them from their counterparts in Minkowski spacetime. Since the
differential operator is defined independently of the existence
of a metric structure on a manifold, we have that $\check{\rm d}
= {\rm d}$; on the other hand, the codifferential operator requires
a metric for its definition and therefore we have $\check{\delta}
\neq \delta$. In order to write the euclidean version of these
operators in Minkowski spacetime the recipe is to replace the
usual Clifford product by the vee product in formulas (\ref{eq.6.7}),
that is,
\begin{equation}
\label{eq.6.8}
\check{\rm d}\Phi = \frac{1}{2}\left( \nabla\vee\Phi +
\hat{\Phi}\vee\overleftarrow{\nabla}\right) , \qquad
\check{\delta}\Phi = \frac{1}{2}\left( \nabla\vee\Phi -
\hat{\Phi}\vee\overleftarrow{\nabla}\right) ,
\end{equation}

The expressions $\nabla\vee\Phi = \gamma^\mu\vee\partial_\mu\Phi$
and $\Phi\vee\overleftarrow{\nabla} = \partial_\mu\Phi\vee\gamma^\mu$
can be calculated as before. We have for $\Phi_k \in \Lambda^k$
that ($i = 1,2,3$)
\begin{equation}
\label{eq.6.9}
\nabla\vee\Phi_k = \gamma^0\partial_0\Phi + (-1)^k(\partial_i\Phi_k)
\gamma^i , \qquad
\Phi_k \vee \overleftarrow{\nabla} =
(-1)^k \gamma^i\partial_i \Phi_k + \partial_0\Phi_k \gamma^0 ,
\end{equation}
and therefore
\begin{equation}
\label{eq.6.10}
\nabla\vee\Phi = \gamma^0\partial_0\Phi + \partial_i\hat{\Phi}\gamma^i ,
\qquad
\Phi\vee\overleftarrow{\Phi} = \gamma^i\partial_i\hat{\Phi} +
\partial_0\Phi\gamma^0 .
\end{equation}

Now, using the expression for $\check{\rm d}$ given in (\ref{eq.6.8})
we see that
\begin{equation}
\label{eq.6.11}
\begin{split}
\check{\rm d}\Phi & = \frac{1}{2}
\left( \gamma^0\partial_0\Phi + \partial_i\hat{\Phi}\gamma^i +
\gamma^i\partial_i\Phi + \partial_0\hat{\Phi}\gamma^0 \right) \\
& = \frac{1}{2}\left(\gamma^\mu\partial_\mu\Phi +
\partial_\mu\hat{\Phi}\gamma^\mu\right) = {\rm d}\Phi ,
\end{split}
\end{equation}
that is, $\check{\rm d} = {\rm d}$ as expected.

On the other hand, using the expression for $\check{\delta}$
given in (\ref{eq.6.9}) we have that
\begin{equation}
\label{eq.6.12}
\begin{split}
\check{\delta}\Phi & = \frac{1}{2}
\left( \gamma^0\partial_0\Phi + \partial_i\hat{\Phi}\gamma^i -
\gamma^i\partial_i\Phi - \partial_0\hat{\Phi}\gamma^0\right) \\
& = \frac{1}{2}\left[ \left(\gamma^0\partial_0\Phi -
\gamma^i \partial_i\Phi \right) - \left(\partial_0\hat{\Phi}\gamma^0
- \partial_i\hat{\Phi}\gamma^i\right) \right] ,
\end{split}
\end{equation}
and clearly $\check{\delta} \neq \delta$, as expected. Using
(\ref{eq.6.6}) for the euclidean Hodge star operator $\star$
written in Minkowski spacetime we have that
\begin{equation}
\label{eq.6.13}
{\rm d}\star \Phi =
\frac{1}{2}\left( -\gamma_5 \gamma^\mu \gamma_0
\partial_\mu\bar{\Phi}\gamma_0 + \gamma_5 \gamma_0
\partial_\mu\tilde{\Phi}\gamma_0\gamma^\mu \right)
\end{equation}
and therefore
\begin{equation}
\label{eq.6.14}
\begin{split}
\star{\rm d}\star \Phi & = \frac{1}{2}\left( \gamma_5 \partial_\mu\Phi
\gamma_0\gamma^\mu \gamma_5\gamma_0 + \gamma_5\gamma_0 \gamma^\mu
\gamma_0\partial_\mu\hat{\Phi}\gamma_5 \right) \\
& = \frac{1}{2}\left(\gamma_0\gamma^\mu\gamma_0\partial_\mu\Phi -
\partial_\mu\hat{\Phi}\gamma_0\gamma^\mu\gamma_0 \right) \\
& = \frac{1}{2}\left[ \left(\gamma^0\partial_0\Phi -
\gamma^i\partial_i\Phi\right) -
\left(\partial_0\hat{\Phi}\gamma^0 - \partial_i\hat{\Phi}\gamma^i \right)
\right] ,
\end{split}
\end{equation}
and comparing this with (\ref{eq.6.8}) we have
\begin{equation}
\label{eq.6.15}
\check{\delta} = \star \check{\rm d}\star = \star {\rm d}\star ,
\end{equation}
as expected.

\subsection{Self-Dual and Anti-Self-Dual Solutions of Gauge Field
Equations}

Since for what follows it is irrelevant whether or not we
are considering abelian or non-abelian gauge fields, we shall
consider the electromagnetic field -- U(1) gauge fields --
as an example for our discussion.

Let us consider the free Maxwell equations in Minkowski
spacetime,
\begin{equation}
\label{eq.6.16}
{\rm d} F = 0 , \qquad \delta F = 0 .
\end{equation}
These equations don't have self-dual and/or anti-self-dual
solutions, that is, solutions satisfying the conditions
$F = \pm \ast F$ in the {\it real\/} case. Such solutions
are possible in Minkowski spacetime only if we {\it complexify\/}
the underlying algebra considering a complex $F$. On the other
hand, self-dual and/or anti-self-dual solutions of Maxwell
equations exist in euclidean spacetime without the need of
any complexification. This can be described in our scheme by
writing the equivalent Maxwell equations in Minkowski
spacetime using the vee product.

The equations
\begin{equation}
\label{eq.6.17}
\check{\rm d} F = 0 , \qquad \check{\delta} F = 0 ,
\end{equation}
are written in terms of the {\it real\/} Clifford algebra
of Minkowski spacetime and admit self-dual and anti-self-dual
solutions satisfying $F = \pm \star F$. This condition
reads
\begin{equation}
\label{eq.6.18}
F = \pm \gamma_5 \gamma_0 \bar{F} \gamma_0 = \pm
\gamma_5 \gamma_0 \tilde{F} \gamma_0 .
\end{equation}
Now writting
\begin{equation}
\label{eq.6.19}
F = E + \gamma_5 B ,
\end{equation}
where
\begin{equation}
\label{eq.6.20}
E = \frac{1}{2}(F - \gamma_0 F \gamma_0) , \qquad
\gamma_5 B = \frac{1}{2}(F + \gamma_0 F \gamma_0) ,
\end{equation}
and such that $ E \gamma_0 = -\gamma_0 E$ and $\gamma_5 B \gamma_0 =
\gamma_0 \gamma_5 B$, we can see that (\ref{eq.6.18})
is satisfied for
\begin{equation}
\label{eq.6.21}
E = \pm B ,
\end{equation}
which are the usual self-dual and anti-self-dual solutions of
the euclidean case but now written in terms of Minkowski spacetime.

\section{Conclusions}

Given the Clifford algebra of a quadratic space with a 
given signature, we have defined a new product in this
structure such that 
it simulates the Clifford product of a quadratic space 
with another arbitrary signature different from the
original one. We have used this in order to give an 
algebraic approach to the so called Wick rotation. 
We have used this new product in order to simulate the
product associated with the Minkowski spacetime in terms
of the Clifford algebra of the euclidean spacetime. 
We have also shown how to write the minkowskian Dirac 
equation in euclidean spacetime and in the other way
how to write the Hodge star operator and
the differential and codifferential operators corresponding
to the euclidean case in terms of Minkowski spacetime,
discussing self-dual and anti-self-dual solutions for
the gauge field equations in this case.

\vspace*{2ex}

\noindent {\bf Acknowledgments:} D.M. is grateful to
Departamento de Matem\'atica Aplicada, Universidade
Estadual de Campinas (UNICAMP), for hospitality and
support during the preparation of this work. J.M.P. acknowledges
support from the Spanish Ministry of Education contract No.
PB96-0384 and the Institut d'Estudis Catalans.

\end{document}